\documentstyle[preprint,amstex,fancyheadings,prb,aps]{revtex}
\begin{document}
\def\bbbone{{\mathchoice {\rm 1\mskip-4mu l} {\rm 1\mskip-4mu l}
{\rm 1\mskip-4.5mu l} {\rm 1\mskip-5mu l}}}
%\preprint{LA-UR-98-????}
%
% \draft command makes pacs numbers print
\draft

\title{Issues and Observations on Applications of
the Constrained-Path Monte Carlo Method to Many-Fermion Systems} 

% repeat the \author\address pair as needed
\author{J.~Carlson, J.E.~Gubernatis, and G. Ortiz}
\address{Theoretical Division,
Los Alamos National Laboratory, Los Alamos, NM 87545}
\author{Shiwei Zhang}
\address{Departments of Physics and Applied Science\\
College of William and Mary, Williamsburg, VA 23187}

\date{\today}
\maketitle

\begin{abstract}
We report several important observations that
underscore the distinctions between the constrained-path Monte Carlo
method and the continuum and lattice versions of the fixed-node method.
The main distinctions stem from the differences in the state
space in which the random walk occurs and in the manner in which the
random walkers are constrained. One consequence is that in the
constrained-path method the so-called mixed estimator for the energy is
not an upper bound to the exact energy, as previously claimed. Several
ways of producing an energy upper bound are given, and relevant 
methodological aspects are illustrated with simple examples.
\end{abstract}

% insert suggested PACS numbers in braces on next line
\bigskip


%\narrowtext
\section{Introduction}
\label{sec1}

It is arguable that the fixed-node \cite{Anderson} and
constrained-path \cite{shiwei1} quantum Monte Carlo methods are the
two most powerful and useful simulation techniques for computing
accurate ground-state ($T= 0$ K) properties of large systems of
interacting quantum particles. As the significantly older method, the
fixed-node method has been well studied, and its properties are well
documented.\cite{reviews}  Less is known about the constrained-path
method, but because of our recent use of this method,
\cite{shiwei2,mariana1,janez1,janez2,mariana2} we now can report 
several important experiences
and observations that underscore features distinguishing it from both
the continuum \cite{moskowitz,reynolds} and lattice
\cite{FNlatt} versions of the fixed-node method.

There is a very strong analogy between the fixed-node and
constrained-path methods. Both, in a sense, are auxiliary-field
methods, both project the ground-state wave function from a trial
wave function by an importance-sampled, branched random walk, and both
place a constraint on this random walk to prevent the fermion sign
problem from rapidly producing exponentially growing variances. A
number of technical details for their implementation are the
same. In fact, the formal development of the constrained-path method
\cite{shiwei1} relied on the existence of the fixed-node method.  The
three principal differences between the methods are (a) the state
space where random walks have their support, (b) the manner by which
random walkers are constrained, and (c) the part of the imaginary-time
propagator that is stochastically sampled.

The continuum version of the fixed-node method works in a first 
quantized representation and operates in coordinate space. The basis 
states are the complete orthonormal set of the particle 
configurations. The constrained-path method works in a second 
quantized representation and operates in Fock space. Its basis states 
are the over-complete non-orthonormal set of Slater determinants
\begin{equation} 
 |\phi\rangle = a_1^\dagger a_2^\dagger\cdots a_N^\dagger|0\rangle \ ,
\end{equation}
where the $a_i^\dagger = \sum_{j=1}^M \Phi_{ij} c_j^\dagger$ creates a
fermion in a quasi particle state $i$ defined relative to $M$ possible single
particle states $j$ created by the operator
$c_j^\dagger$, and $| 0 \rangle$ represents the vacuum. (In the
one-band Hubbard model $M$ will be the number of lattice sites.) In this basis
classes of many-electron wave functions like the BCS wave function are
more easily used, and many-particle expectation values like
superconducting pairing correlation functions are more easily
evaluated than possible with the fixed-node method.\cite{shiwei1} The
ease in the evaluation of ground-state observables, for example, is a
consequence of the ease in evaluating single-particle propagators and
using Wick's theorem \cite{negele} to express any multi-particle
propagator as a linear combination of products of one-particle
propagators.

The differences in bases generate a difference in the way the random
walks are constrained. Both methods rely on a trial state $|\Psi_T 
\rangle$ to perform the constraint. In the continuum version 
of the fixed-node method with ${\bf r}_i$ representing a particle's
position, the random walks are confined within a 
surface defined by $\langle{\bf R}|\Psi_T\rangle = \Psi_T({\bf R}) 
\equiv \Psi_T({\bf r}_1,{\bf r}_2,\dots,{\bf r}_N) > 0$, whereas in 
the constrained-path method, only random walkers $|\phi\rangle$ 
satisfying $\langle\Psi_T|\phi\rangle > 0$ are permitted. The 
fixed-node method solves Schr\"odinger's equation for the ground-state 
wave function inside the nodal surface. Unless that surface is exact, 
only an approximate solution is obtainable. The constrained-path 
condition, as we will discuss, has different implications. In certain
cases, including some simple examples detailed below,
the constraint is never invoked and hence
the constrained-path method can sometimes 
produce the exact solution even for systems of interacting fermions.
We also give examples where the solution, though approximate,
is extremely accurate.  These examples include a closed-shell
Hubbard model with a large positive $U$.

The resulting stochastic dynamics in the basis space (both coordinate
and Slater determinantal manifolds) is a Markov process generated by a
conditional probability connected to the imaginary-time $\tau$
propagator $\exp[ -\tau H]$, where $H=T+V$ is the Hamiltonian
representing the system and, as usual, $T$ and $V$ are the kinetic and 
potential energy operators, respectively. The kinetic energy 
propagator is non-diagonal in the coordinate basis representation and 
its action can be viewed as a diffusion process in the basis space. On 
the other hand, in the Slater determinant representation it is the 
potential energy kernel which, after a Hubbard-Stratonovich 
transformation, generates the Markov chain. As a consequence, 
non-interacting problems do not suffer from the fermion sign 
instability.

The main purpose of the paper is to illustrate that the
constrained-path method {\bf is not} equivalent to the 
fixed-node method in the space 
of Slater determinants. In Section~\ref{sec2} we will summarize the 
essential mathematical structure of both methods. In 
Section~\ref{sec3}, we discuss the mixed estimator for the energy and 
the extent to which it is an upper bound on the ground-state energy. In
particular, we will conclude and illustrate that in the 
constrained-path method it is not, in general, an upper bound, 
contrary to previous claims and to the fixed-node method.
In Section~\ref{sec4} we provide a correction to the mixed 
estimator that makes it a rigorous upper bound, plus several 
alternative ways to produce energy estimates that are upper bounds. In 
Section~\ref{sec5}, we conclude by commenting on areas needing 
additional clarification and several other differences between the 
methods. Some of these differences will be illustrated in the Appendix 
where we present a constrained-path simulation on a small toy problem 
for which many of the details can be generated analytically. Of 
particular emphasis here will be the effects of matrix stabilization.

\section{Summary of the Two Methods}
\label{sec2}

Both the fixed-node and constrained-path methods project the 
ground state $|\Psi_0\rangle$ from the long-time solution of the 
imaginary-time $\tau$ representation of Schr\"odinger's equation 
specified by a Hamiltonian $H$
\begin{equation}
 \frac{\partial|\Psi\rangle}{\partial\tau} = - (H-E_0) |\Psi\rangle \ .
\end{equation}
Provided $N_0=\langle\Psi_0|\Psi(0)\rangle\not=0$ and $H$ is 
time-independent, the formal solution
\begin{equation}
  |\Psi(\tau)\rangle = e^{-\tau (H-E_0)}|\Psi(0)\rangle
\end{equation}
has the property
\begin{equation}
  \lim_{\tau \rightarrow \infty} |\Psi(\tau)\rangle 
     = N_0|\Psi_0\rangle \ .
\end{equation}
On the computer this limit is accomplished iteratively
\begin{equation}
  |\Psi'\rangle = e^{-\Delta\tau(H-E_T)}|\Psi\rangle \ ,
\end{equation}
where $\Delta \tau $ is a small number, $n \Delta \tau = \tau$,
$n$ is the current number of iterations (often called time slices), and
$E_T$ is a trial guess at the ground-state energy $E_0$.  The
iterative process is converted into a stochastic sampling process. As
the matrix elements of the propagator $\exp [ - \tau (H-E_T)]$ between
different anti-symmetric wave functions are not always positive definite,
constraints in the sampling are necessary to insure this.
Additionally, importance sampling is also required to control the
variances of computed results.  If $E_T$ is adjusted so that it equals
$E_0$, then as $\tau\rightarrow\infty$, the iteration becomes
stationary, i.e. $\partial|\Psi\rangle/\partial\tau =0$ and
$|\Psi\rangle \propto |\Psi_0\rangle$.

For simplicity, we will exclude branching from our discussions and
consider only time-reversal symmetric Hamiltonians, that is, real
symmetric operators for which the ground-state wave functions can
always be chosen to 
be real. This analysis leaves out the very important case of systems
in the presence of external magnetic fields.\cite{ocm} Here we
compare the constrained-path method to the continuum fixed-node
approach.  There are some technical differences between the continuum
fixed-node method \cite{moskowitz,reynolds} and the lattice version
\cite{FNlatt} we prefer to omit. In this regard we comment 
that the constrained-path method does not distinguish between lattice
and continuum fermions: both are treated on an equal footing.

\subsection{Fixed-Node method}

In the fixed-node method, one represents the ground state as
$|\Psi_0\rangle = \sum_{{\bf R}}|{\bf R}\rangle\langle {\bf
R}|\Psi_0\rangle = \sum_{{\bf R}} \Psi_0({\bf R})|{\bf R}\rangle$ where
$\Psi_0({\bf R})>0$. Asymptotically, the Monte Carlo procedure samples 
from the distribution $P({\bf R}) = \Psi_0({\bf R})/\sum_{{\bf R}} 
\Psi_0({\bf R})$.

In the fixed-node method, one projects the iteration onto the basis of 
particle configurations $\{|{\bf R}\rangle\}$
\begin{equation}
  |\Psi'\rangle = e^{-\Delta\tau(H-E_T)}|\Psi\rangle \ .
\label{eq:onestep}
\end{equation}
Projecting this equation onto $\langle{\bf R}'|$ and inserting
$\sum_{{\bf R}}|{\bf R}\rangle\langle{\bf R}| = 1$ leads to
\begin{equation}
 \langle{\bf R}'|\Psi'\rangle = \Psi'({\bf R}') =
  \sum_{\bf R}\, \langle{\bf
R}'|e^{-\Delta\tau(H-E_T)}|{\bf R}\rangle   \Psi({\bf R}) \ ,
\end{equation}
and correspondingly the imaginary-time Schr\"odinger's equation becomes
\begin{equation}
 -\frac{\partial\Psi({\bf R},\tau)}{\partial\tau}
               = [-D\nabla^2+V({\bf R})-E_T]\Psi({\bf R},\tau)
\end{equation}
where $D=\hbar^2/2m$, $m$ is the fermion mass, and $V({\bf R})$ is the
potential energy.

$\Psi({\bf R})$ must be positive to be interpreted as the limiting
probability distribution of the Markov chain. Fixing the node forces
this by prohibiting any change in the particle configuration ${\bf R}
\rightarrow {\bf R}'$ that changes the sign of $\Psi({\bf R})$.
We will denote the ground state produced under the constraint, i.e.
under the fixed-node condition, as $|\Psi_c\rangle$ and the eigenvalue
of the propagation (Eq. \ref{eq:onestep}) as $\exp [ - \Delta
\tau(E_g - E_T)]$.  This eigenvalue defines $E_g$, the
growth energy.

After the so-called short-time approximation is made on the kernel of
the integral, which is equivalent to making a Trotter approximation
and a Hubbard-Stratonovich transformation on the exponential of the
kinetic energy,\cite{shiwei1} the positivity of $\langle{\bf R}'|\exp[-\Delta\tau 
(H-E_T)]|{\bf R}\rangle $ is trivially satisfied so it can be 
interpreted as a transition probability defining a Markov chain. We 
will call the resulting approximation $K({\bf R}\rightarrow{\bf R}')$.

It is critical to importance sample in order to reduce statistical
fluctuations, especially when the potential $V({\bf R})$ has some
singularities, for example, like the $1/r$ Coulomb singularity. This
means we generate a new distribution 
$\tilde\Psi_c({\bf R})\equiv \Psi_T({\bf R})\Psi_c({\bf R})$ satisfying
\begin{equation}
 \tilde\Psi_c'({\bf R}') = \sum_{\bf R}\, \tilde K({\bf R} 
\rightarrow{\bf R}') \tilde\Psi_c({\bf R}) \ ,
\end{equation}
where $\tilde K({\bf R}\rightarrow{\bf R}') =
\Psi_T({\bf R}')K({\bf R}\rightarrow{\bf R}')/\Psi_T({\bf R})$. The
new configurations are now sampled with a different probability. The
new distribution also satisfies a different equation of motion
\begin{equation}
 -\frac{\partial\tilde\Psi_c({\bf R},\tau)}{\partial\tau}
            = -D\nabla^2\tilde\Psi_c({\bf R},\tau)
              +D\nabla\cdot[\tilde\Psi_c({\bf R},\tau){\bf F}({\bf 
              R})] +(E_L({\bf R})-E_T) \tilde\Psi_c({\bf R},\tau) \ ,
\end{equation}
where the ``quantum drift'' ${\bf F} =2\nabla\ln \Psi_T $ and the
``local energy'' $E_L({\bf R}) = H\Psi_T({\bf R})/\Psi_T({\bf R})$.

The Monte Carlo procedure represents the multi-dimensional integral as
a set of random walkers $\{ |{\bf R}\rangle \}$ where each member of 
the set is a different allowed particle configuration. A new 
configuration $|{\bf R}'\rangle$ is sampled from $\tilde K({\bf R}
\rightarrow {\bf R}')$ and rejected, thereby terminating this random 
walker, if the resulting value of the wave function is negative. 

Since $\Psi_c({\bf R},\tau)=\tilde\Psi_c({\bf R},\tau)/\Psi_T({\bf
R})$, a variational upper bound to the true energy is
\begin{equation}
 E_v = \frac{\sum_{\bf R}\, \Psi_cH\Psi_c}
            {\sum_{\bf R}\, \Psi_c^2}
         \ge E_0 \ .
\end{equation}
At large times walkers are distributed with a probability density
$\Psi_T(R) \Psi_c(R)$, and both $\Psi_T$ and $\Psi_c$ go to zero
linearly near the nodal surface.
Since the Hamiltonian and the constraint are all local operators,
this implies  that the growth energy $E_g$ is
equal to the mixed estimate of the energy $E_m$ 
\cite{moskowitz,reynolds}
\begin{equation}
E_g = E_m   \equiv  \frac{ \langle \Psi_T | H | \Psi_c \rangle}
{ \langle \Psi_T | \Psi_c \rangle}  
 =  \frac{\sum_{\bf R}\, \Psi_cH\Psi_T}{\sum_{\bf R}\,\Psi_c\Psi_T }
 =  \frac{\sum_{\bf R}\, \tilde\Psi_c({\bf R},\tau) 
          E_L({\bf R})}{\sum_{\bf R}\, \tilde\Psi_c({\bf R},\tau)} \ .
\label{eq:eestimates}
\end{equation}
The constrained propagator is identical to the exact one except near
the nodal surface where the constraint acts.  The constraint discards
contributions that are orthogonal to both $\Psi_T$ and $\Psi_c$, and
hence this region gives no net contribution to either $\langle \Psi_T
| H | \Psi_c \rangle$ or $\langle \Psi_c | H | \Psi_c
\rangle$. Therefore, the variational estimate of the energy $E_v$ is
identical to $E_g$ and $E_m$, and all are variational upper bounds.
$E_m$ is more easily, accurately, and efficiently computed than 
$E_g$ or $E_v$.

Several characteristics of the fixed-node method are: 1) The nodal
surface of $\Psi_c({\bf R})$ is exactly the same as that of
$\Psi_T({\bf R})$;  2) The exact 
ground-state energy is obtained {\it only if} the nodal surface of 
$\Psi_T({\bf R})$ is exact;  and 3) even for the trivial 
case of $V({\bf R})=0$, unless the exact nodal surface is used, only 
an approximate solution is produced. 

\subsection{Constrained-Path Method}

In the constrained-path method, one represents the ground state as
$|\Psi_0\rangle = \sum_\phi c_\phi |\phi\rangle$ where the Slater
determinants $|\phi\rangle$ are chosen so that all $c_\phi
>0$. Asymptotically, the Monte Carlo procedure samples from the
distribution $\pi(\phi) = c_\phi/\sum_\phi c_\phi$. The decomposition
of $|\Psi_0\rangle$ in terms of the $|\phi\rangle$'s is not unique. One
could just a well have $|\Psi_0\rangle = \sum_\phi d_\phi
|\phi\rangle$ where $d_\phi >0$. We will simply write $|\Psi_0\rangle
= \sum_\phi |\phi\rangle$.

The constrained-path method works in a basis of Slater
determinants. Again
one iterates Eq. \ref{eq:onestep}, placing constraints on the
random walks.
A different kind of constraint is needed because a different
basis is used.
Here the sign problem is caused by transitions
from a region where the overlap
$\langle \Psi_T | \phi \rangle$
is positive to a region where it is negative. These two regions are 
not physically distinguishable; they involve merely the
exchange of fermions.  Hence an arbitrary wave function can always be
expanded in the restricted bases where 
$\langle \Psi_T | \phi \rangle$
is purely positive or purely negative.  The original 
propagation mixes these two degenerate bases indiscriminately, causing
a sign problem.

To break this plus-minus symmetry, the random walks are constrained
to the region $\langle \Psi_T | \phi \rangle > 0.$
This is an approximation because in general a wave function will have 
both positive and negative coefficients $c_\phi$
when expressed in this basis.  However, the constrained propagation
yields all the $c_\phi > 0$. 
To compare with the fixed-node method,
we sketch some additional
details: After the application of a Trotter approximation and
Hubbard-Stratonovich transformation, the iterative equation becomes
\begin{equation}
 |\Psi'\rangle = \sum_{\bf x}\, P({\bf x}) B({\bf x})|\Psi\rangle \ ,
\end{equation}
where ${\bf x}$ (the Hubbard-Stratonovich field) is to be interpreted 
as a 
multi-dimensional random variable distributed according to $P({\bf 
x})$, and $B({\bf x})$ is an operator approximating $\exp[-\Delta\tau 
H]$ for a given value of the random variable, whose general structure
is a product of exponentials of one-body operators. $B({\bf x})$ has 
the property of transforming one Slater determinant into another. The 
Monte Carlo method is used to evaluate the multi-dimensional 
integration by using multiple random walkers $|\phi\rangle$, and for 
each walker, sampling a ${\bf x}$ from $P({\bf x})$ and then 
generating a new walker
\begin{equation}
 |\phi'\rangle = B({\bf x})|\phi\rangle \ .
\end{equation}
Thus, if $\langle\Psi_T|\phi'\rangle >0$, $\langle\Psi_T|B({\bf
x})|\phi\rangle >0$.

Since the basis of Slater determinants is non-orthogonal
and over-complete, each member of the basis, in general, is a linear
combination of the others, i.e., $|\phi\rangle =
\sum_{\phi'}'a_{\phi'}|\phi'\rangle$. A prime is on the
summation symbol because while the summation may be over an infinite
number of Slater determinants, the ones used need not exhaust the
basis. While $|\Psi_T\rangle$ may constrain a $|\phi\rangle$ to be in
the ``positive'' set, this $|\phi\rangle$ can overlap with a state in
the ``negative'' set. In contrast to the fixed-node condition, the
constrained-path condition does not separate the basis into orthogonal
sets. Whereas the fixed-node condition must produce an approximate
solution unless the nodes are exact, the constrained-path method
can sometimes produce the exact solution even if the constraining wave 
function is approximate and has the wrong nodal surface in
configuration space.

Importance sampling is also implemented in the constrained-path method.
With $|\tilde\phi\rangle = \langle\Psi_T|\phi\rangle|\phi\rangle$
the iteration on each walker becomes
\begin{equation}
 |\tilde\phi'\rangle = B({\bf x})|\tilde\phi\rangle \ ,
\end{equation}
but now the random variable ${\bf x}$ is sampled from $\tilde P({\bf
x}) \propto \langle\Psi_T|\phi'\rangle P({\bf
x})/\langle\Psi_T|\phi\rangle$. We have $|\tilde\Psi\rangle =
\sum_\phi \langle\Psi_T|\phi\rangle|\phi\rangle$.

Again a variational estimate of the energy $E_v$ can
be constructed from $|\Psi_c\rangle$
\begin{equation}
 E_v=\frac{\langle\Psi_c|H|\Psi_c\rangle}
      {\langle\Psi_c|\Psi_c\rangle} = E_g
             \ge E_0 \ ,
\end{equation}
but now the connection among $E_v$, $E_g$ and $E_m$ is unclear 
because the constraint discards configurations which are 
orthogonal to $|\Psi_T\rangle$ and these discarded configurations
are not necessarily orthogonal to $|\Psi_c\rangle$.
As we will argue
in the next section, the mixed estimator is not always an upper bound 
to $E_0$. This retracts previous claims of $E_m$ being an upper
bound.\cite{shiwei1} 

Several characteristics of the constrained-path method are: 1) The
nodal surface of $\langle \phi|\Psi_c\rangle$ is not the same as that 
of $\langle \phi|\Psi_T\rangle$; 2) In some cases, the exact 
ground-state energy can be obtained {\it even if} the nodal surface of 
$\langle \phi|\Psi_T\rangle$ is approximate; and 3) For the trivial 
case of $V({\bf R})=0$, the exact solution is produced. 

Perhaps the best known examples demonstrating the second point are the
half-filled positive-$U$ Hubbard models and negative-$U$ Hubbard
models, two classes of models that do not have a sign problem. To
illustrate the first two characteristics of the CPMC 
method, we consider the following half-filled positive-$U$ Hubbard model
\begin{equation}
H = -t \sum_{\sigma=\uparrow,\downarrow} \left( c^{\dagger}_{1,\sigma} 
c^{\;}_{2,\sigma} + c^{\dagger}_{2,\sigma} c^{\;}_{1,\sigma} \right) + 
U \sum_{i=1}^2 n_{i \uparrow} n_{i \downarrow}\ , 
\end{equation}
which is also a simple model for a Heitler-London molecule. The
two-particle ground state is given by
\begin{equation}
| \Psi_0 \rangle = \frac{1}{\sqrt{2 \tilde{t}^2 + (U - E_0)^2}}  
\left[ \tilde{t} \left( c^{\dagger}_{1 \uparrow} c^{\dagger}_{1 
\downarrow} + c^{\dagger}_{2 \uparrow} c^{\dagger}_{2 \downarrow} 
\right) + \frac{U - E_0}{\sqrt{2}} \left( c^{\dagger}_{1 \uparrow} 
c^{\dagger}_{2 \downarrow} + c^{\dagger}_{2 \uparrow} c^{\dagger}_{1 
\downarrow} \right) \right] | 0 \rangle \ ,
\end{equation}
where $\tilde{t} = \sqrt{2} t$, and the ground-state energy is $E_0 =
U/2 - \sqrt{(U/2)^2 + 2 \tilde{t}^2}$. This state cannot be
represented by a single Slater determinant, unless $U=0$. (See Eq. 1.)

Since we want to study the
nodal structure of different states, we need to parameterize the
differentiable manifold of Slater determinants of two
particles. We choose coordinates such that a
generic point in the manifold $(\theta_1,\theta_2)$ corresponds to the
normalized Slater determinant
\begin{equation}
| \phi \rangle = \left( \cos \theta_1 \ c^{\dagger}_{1 \uparrow} + 
\sin \theta_1 \ c^{\dagger}_{2 \uparrow} \right) \left( \cos \theta_2 
\ c^{\dagger}_{1 \downarrow} + \sin \theta_2 \ c^{\dagger}_{2 
\downarrow} \right) | 0 \rangle \ .
\end{equation}
Alternatively, from Eq. 1, we can represent this state by the product
of two $2\times 1$ matrices, i.e., $\Phi =
\Phi_\uparrow\Phi_\downarrow$ where
\begin{equation}
 \Phi_\uparrow = \left(\begin{array}{c}
         \cos\theta_1 \\
         \sin\theta_1 
         \end{array}\right)
\end{equation}
and
\begin{equation}
 \Phi_\downarrow = \left(\begin{array}{c}
         \cos\theta_2 \\
         \sin\theta_2 
         \end{array}\right)
\end{equation}
Then, 
\begin{equation}
\langle \phi | \Psi_0 \rangle = \frac{1}{\sqrt{2 \tilde{t}^2 + (U - 
E_0)^2}} \left[ \tilde{t} \cos(\theta_1-\theta_2) + \frac{U - 
E_0}{\sqrt{2}} \sin(\theta_1+\theta_2) \right] \ .
\end{equation}
In Fig.~\ref{fig1} we display contour plots of this function for 
different values of $U$.  
Clearly the nodal surfaces of $\langle \phi | \Psi_0 \rangle$ are 
different for the various values of $U$. Nevertheless, in the absence
of importance sampling, one can prove analytically that for any $U$,
$\langle \phi | \Psi_T  
\rangle = \langle\phi|\Psi_0(U=0)\rangle$ remains positive during the
whole imaginary-time evolution;  
that is, the nodal constraint is never invoked, and therefore, the 
exact solution is obtained after a large-$\tau$ projection with the
result that the nodal surfaces of $\langle \phi| \Psi_c \rangle$ and 
$\langle \phi| \Psi_0\rangle$ are the same but different from
$\langle\phi|\Psi_T\rangle$.

\section{Mixed Estimator of the Energy}
\label{sec3}

Independent of a quantum Monte Carlo process, when is the
mixed estimator for the energy  (Eq. \ref{eq:eestimates})
an upper bound for the ground-state energy or even the exact value?
Three cases are apparent:
\begin{enumerate}
\item If $| \Psi_T \rangle \equiv | \Psi_c \rangle$, then $E_m \geq 
E_0$ \ .
\item If $| \Psi_T \rangle \equiv | \Psi_0 \rangle$, or $| \Psi_c 
\rangle \equiv | \Psi_0 \rangle$, then $E_m = E_0$ \ .
\item If $| \Psi_c \rangle = {\rm U}^{2 n} | \Psi_T \rangle$, where $n$ is an
integer, ${\rm U}$ is a Hermitian operator, and $\left[H,{\rm U}\right]=0$, then
$E_m \geq E_0$ \ .
\end{enumerate}
Case 1 is simply the Rayleigh-Ritz variational principle.
Case 2 is perhaps the most important feature of the mixed estimator:
a good approximation to the ground-state wave function will produce a
good approximation to the ground-state energy. Case 3 is what happens
in quantum Monte Carlo simulations: in principle,
with ${\rm U}=\exp[-\Delta\tau H]$, an upper bound on the energy is
automatically produced. In the simulations, ${\rm U}^{2n}\rightarrow 
{\rm U}_{2n}\cdots {\rm U}_2{\rm U}_1$ where ${\rm U}_i=\exp[-\Delta\tau H_i]$  with $H_i$ 
representing an effective Hamiltonian satisfying $\left[H,{\rm U}_i\right]
\not=0$, so in general $E_m$ is not a rigorous bound. But since 
$\left[H,{\rm U}_i\right]\approx 0$, $E_m$ is in general expected to be 
a good estimate and a bound.
Clearly to the extent that the constrained-path method in principle can
produce the exact state vector, the mixed estimate of energy can be
exact. 

On general grounds we can say that if our constrained evolution 
defines a Markov process with a stationary distribution $
| \Psi_c \rangle$, such that 
\begin{equation}
H_{\rm eff} | \Psi_c \rangle = E_g | \Psi_c \rangle \ , 
\end{equation}
and $H_{\rm eff} = H + \delta H$, then 
\begin{equation}
\frac{\langle \Psi_c | H| \Psi_c \rangle}{\langle \Psi_c 
| \Psi_c \rangle} = E_m + \delta E_m \geq E_0 \ , 
\end{equation}
where 
\begin{equation}
\delta E_m = \frac{\langle \Psi_T | \delta H| \Psi_c 
\rangle}{\langle \Psi_T | \Psi_c \rangle} - \frac{\langle \Psi_c | 
\delta H |\Psi_c \rangle}{\langle \Psi_c | \Psi_c \rangle} \ .
\end{equation}
It is clear that if $\delta E_m \leq 0$, $E_m$ is an upper bound to 
the ground-state energy. However, in general, this is not necessarily 
the case. 

It is interesting to mention that in the usual fixed-node approach,
where the state space manifold is the coordinate space, $\delta H$
represents a hard-wall potential, i.e. it is infinite on the set of
configurations $\{|{\bf R}_T\rangle\}$ defined by $\langle {\bf R}_T | 
\Psi_T \rangle=0$. Then $E_m$ is an upper bound to $E_0$. We can see 
this by minimizing the following constrained functional
\begin{eqnarray}
 \lefteqn{F\Bigl[|\Psi_c\rangle, \langle\Psi_c|;\eta_{{\bf R}_T},
  \eta_{{\bf R}_T}^*\Bigr] = } \nonumber\\
  & & \langle\Psi_c|H|\Psi_c\rangle - E\langle\Psi_c|\Psi_c\rangle
  - \sum_{{\bf R}_T} \eta_{{\bf R}_T}^* \langle {\bf R}_T|\Psi_c\rangle
  - \sum_{{\bf R}_T} \eta_{{\bf R}_T}   \langle\Psi_c|{\bf R}_T
\rangle \ ,
\end{eqnarray}
where $\langle {\bf R}_T|\Psi_T\rangle = \Psi_T({\bf R}_T) =0 $
defines the nodal surface ${\cal N}_T$.  The resulting Euler's
equations are
\begin{eqnarray}
  H |\Psi_c\rangle &=& E |\Psi_c\rangle 
         + \sum_{{\bf R}_T} \eta_{{\bf R}_T} |{\bf R}_T \rangle \ ,\\
  \langle\Psi_c|H &=& \langle\Psi_c|E 
         + \sum_{{\bf R}_T} \eta_{{\bf R}_T}^* \langle {\bf R}_T| \ ,\\
  \Psi_c({\bf R}_T) &=& 0 \ ,
\end{eqnarray}
which lead to
\begin{equation}
 E = \frac{\langle \Psi_T|H|\Psi_c\rangle}
           {\langle \Psi_T|\Psi_c\rangle} 
    = \frac{\langle \Psi_c|H|\Psi_c\rangle}
           {\langle \Psi_c|\Psi_c\rangle} \ge E_0 \ .
\end{equation}
From the first equation
\begin{equation}
 \langle{\bf R}|H|\Psi_c\rangle = E \ \langle {\bf R}|\Psi_c\rangle 
   + \sum_{{\bf R}_T} \eta_{{\bf R}_T} \langle{\bf R}|{\bf R}_T
     \rangle \ ,
\label{eq30}
\end{equation}
\begin{equation}
 \sum_{{\bf R}'}\langle{\bf R}|H|{\bf R}'\rangle
       \langle{\bf R}'|\Psi_c\rangle = 
E \ \Psi_c({\bf R})+\sum_{{\bf R}_T}\eta_{{\bf R}_T}\delta_{{\bf
RR}_T} \ ,
\label{eq31}
\end{equation}
\begin{equation}
 H({\bf R}) \ \Psi_c({\bf R}) = E \ \Psi_c({\bf R})
 + \sum_{{\bf R}_T}\eta_{{\bf R}_T}\delta_{{\bf RR}_T} \ .
\label{eq32}
\end{equation}  
Solving the constrained (fixed-node) problem is equivalent to solving 
$H({\bf R}) \ \Psi_c({\bf R}) = E \ \Psi_c({\bf R})$ within the
region where $\Psi_T({\bf R})$ has a definite sign, with the boundary
condition $\Psi_c({\bf R}_T) = 0$. In this way $\Psi_c({\bf R})$ is a
continuous function of ${\bf R}$ with discontinuous derivative at
${\bf R} = {\bf R}_T$.

If we try to minimize a similar functional, but we use the 
representation of Slater determinants $|\phi \rangle$, then an 
extremely non-local term, which is not easily handled, appears in the 
resulting Euler equation
\begin{equation}
 \sum_{{\phi}'}\langle \phi |H| \phi' \rangle \ \Psi_c[\phi']
= E \ \Psi_c[\phi'] +\sum_{{\phi}_T}\eta_{{\phi}_T} 
\langle \phi | \phi_T \rangle \ ,
\end{equation}
with $\Psi_T[\phi_T] = \langle\phi_T|\Psi_T\rangle= 0$ but in general $\langle
\phi | \phi_T \rangle \not= \delta_{\phi\phi_T}$.  As before,
one can easily prove that $E_m\geq E_0$. In other words, if we had
used the exact equivalent of the fixed-node constraint, we would have
gotten a variational upper bound using $E_m$. It is important to
stress that the constrained-path condition is a kind of global
constraint as opposed to the local one that represents the fixed-node
constraint in that the constrained-path condition does not impose on
$\langle \phi|\Psi_c\rangle$ the same nodal hypersurface as $\langle
\phi|\Psi_T\rangle$. In fact, we have numerical examples where 
$\langle \phi|\Psi_T\rangle$ does not define the exact nodal 
structure, nevertheless we get the exact ground-state energy for $H$, 
i.e. $|\Psi_c \rangle = |\Psi_0 \rangle$. (See, for instance, the 
example shown at the end of Section \ref{sec2}.)

\section{Energy Estimators Bounding the Ground-State Energy}
\label{sec4}

\subsection{Energy Bounds}

It is possible to construct a variety of other estimators that
produce upper bounds to the ground-state energy.  
In the following we assume we have a state $| \Psi_c \rangle$
that is an eigenstate of the constrained propagator:
\begin{equation}
| \Psi_c \rangle = \lim_{\tau \rightarrow \infty} 
e^{ -\stackrel{\longleftarrow}{\tau H}} \ | \Psi_T \rangle
\end{equation}
with eigenvalue $\exp[- \tau E_g]$
\begin{equation}
e^{-\stackrel{\longleftarrow}{\Delta\tau H}} \ | \Psi_c \rangle =
e^{-\Delta\tau E_g} \ | \Psi_c \rangle \ .
\end{equation}
The arrow indicates the direction of propagation with
the constraint applied to the wave function.  Only those auxiliary
fields that retain a positive overlap with the trial function are
retained in the sampling.  The CPMC paths are not reversible in the
standard sense, and hence the ``time arrow'' of the path is
significant.  In contrast, we denote the original unconstrained
propagator as $\exp [ -\tau H ]$, and since $H$ is Hermitian, the full
propagation is reversible, at least when averaged over paths.  The
effect of the constraint is simply given by the difference between
$\exp[- \stackrel{\longleftarrow}{\tau H}]$ and $\exp[-\tau H]$.

The standard variational upper bound is given by 
\begin{equation}
E_v = \frac{\langle \Psi_c | H | \Psi_c \rangle}
           {\langle \Psi_c |  \Psi_c \rangle}\ge E_0 \ ,
\label{eq37}
\end{equation}
where the function $\langle \Psi_c |$ is the dual state of
$| \Psi_c \rangle$; that is, the constraint is applied in the
opposite $\tau$ direction.
It is possible to calculate $E_v$ directly, for example, by
propagating two populations of random walkers.  These two populations
can be used as independent samples of $\langle \Psi_c |$ and of $|
\Psi_c \rangle$.  Since these walkers should be
independently evaluated (at least prior to the introduction of
importance sampling), we label them as $\langle \Psi_{lc} |$ and $ |
\Psi_{rc} \rangle $.  The importance function will
presumably have to be a function of all the relevant overlaps,
\begin{equation}
I = I ( | \langle \Psi_{lc} | \Psi_{rc} \rangle |,
\langle \Psi_T | \Psi_{rc} \rangle,
\langle \Psi_{lc} | \Psi_{T} \rangle ) \ .
\end{equation}
The overlap of the left and right wave functions may be
negative, so we have to assume the importance function is only
a function of the magnitude of that overlap.  In the absence of
importance sampling, the denominator in Eq. \ref{eq37} is
the sum of the overlap between these two wave functions. Hence
this term should be large in the importance function.

One can also evaluate the energy difference $E_d \equiv E_v$ - $E_g$, 
given by:
\begin{eqnarray}
e^{-\Delta\tau E_g} - e^{- \Delta\tau E_v }
& \approx & \Delta\tau \ E_d \nonumber \\
& = & \frac
{\langle \Psi_c | 
e^{-\stackrel{\longleftarrow}{\Delta\tau H}} -
e^{- \Delta\tau H} 
| \Psi_c \rangle}
{\langle \Psi_c |  \Psi_c \rangle} \ .
\label{eq:evdiff}
\end{eqnarray}
The numerator in this expression is the result of the constraint:
it is simply the overlap of $\langle
\Psi_c |$ with the state $| \Psi_d \rangle$ representing
the difference between the full and constrained propagation: 
\begin{equation}
| \Psi_d \rangle = \left[
e^{-\stackrel{\longleftarrow}{\Delta\tau H}} -
e^{- \Delta\tau H} \right]
| \Psi_c \rangle \ .
\end{equation}
This difference is simply the set of the configurations discarded
via the constraint.  $E_d$ is
zero if these discarded configurations are, on average, orthogonal to
$| \Psi_c \rangle $.  In the fixed-node method the configurations
thrown away are by definition orthogonal both to $|\Psi_T\rangle$ and
$|\Psi_c\rangle$.  Here, though, our configurations are in general
orthogonal only to $|\Psi_T\rangle$, and hence the variational and
mixed estimates of the energy need not be equal.

It is still true, however, that $E_m$ and
$E_g$ are equal in the limit of zero time step.  The density of
configurations near the surface $\langle \Psi_T | \phi \rangle $
goes to zero rapidly so that the surface contributions
to the constrained propagator do not
give a finite contribution to the growth estimate of the energy.

One can evaluate $E_d$ directly.  To evaluate the energy difference we
again need independent right- and left-hand wave functions
$|\Psi_{rc}\rangle$ and $\langle\Psi_{lc}|$.  Dividing the population
into two independent halves representing the left- and right-hand
states, we can evaluate the numerator by taking the overlaps of what
is discarded (the difference between the full and constrained
propagators acting on $| \Psi_c \rangle$) with the independent
solution $\langle \Psi_c |$.  The denominator is just the overlap
$\langle \Psi_c | \Psi_c \rangle$ of the two solutions.  For larger
systems it will likely have high statistical errors, but the numerator
may be small enough that this does not matter. An explicit numerical
example is presented in the next section.

There are several other ways to produce an energy difference $E_d= 0$.
One possibility is to introduce a parameter in the trial state
and vary it until $E_d=0$. Another possibility is changing the
constraint. For example, we could discard configurations $| \phi
\rangle$ for which the normalized overlap with the trial wave function
is less than or equal to some constant $\alpha$:
\begin{equation}
\frac{\langle \Psi_T | \phi \rangle}
{  [ \langle \Psi_T | \Psi_T \rangle 
 \langle \phi | \phi \rangle ]^{1/2} } \leq \alpha \ .
\label{eq:alterconstraint}
\end{equation}
Varying $\alpha$ until the average overlap of the discarded
configurations and the constrained solution $|\Psi_c\rangle$ is
greater than or equal to zero produces a variational upper bound for
the energy, since then $E_g \geq E_v$.  In this case it is not 
necessary to evaluate the denominator of Eq. \ref{eq:evdiff}. Also, 
this procedure is exact for an exact constraining state, since in that 
case we could set $\alpha = 0$.  

 We note that for $\alpha \neq 0$ the mixed estimate $E_m$ is not
in general equal to the growth estimate $E_g$, as there is a finite
surface term that contributes to the difference.  In fact, the
difference $E_g - E_m$ provides a measure of the error introduced
by the constraint.  Numerical examples are provided in the
following section.

This method is general in that it produces a
variational upper bound to the energy for any Hamiltonian
and any constraint.  The only restriction is that $| \Psi_d
\rangle$ has a positive overlap with the eigenstate $\langle
\Psi_c|$ of the constrained propagation.  This restriction naturally
implies a repulsive contribution to $E_g$ and an increase in the value
of the energy.  This algorithm can be made quite general and applied
to a variety of interesting situations.

\subsection{Numerical Example}

In this sub-section we consider a simple numerical example illustrating
the behavior of the various energy estimators.  The particular example
is the 2D Hubbard model on a $4\times 4$ lattice with 5 up spin and 5
down spin electrons.  The exact ground-state energy of this small
system was obtained by direct diagonalization.  For the intermediate
coupling of  $U=8t$ and $t=1$, the energy is -17.51037.\cite{example}

We used the free-particle wave functions for both the constraint and the
importance function in a series of CPMC calculations.  As a
variational wave function, the free-particle wave functions are quite
inaccurate, yielding an energy of -11.50.  
We also used population sizes of 1000 to 2500 configurations,
divided into two halves for independent left- and right-hand
wave functions.  Averages were computed over 30-100 blocks with a 
propagation time (number of steps times $\Delta\tau$) of 2 to 10 per block.
We verified that we have reached the equilibrium state before 
computing averages and that the blocks were large enough to avoid
difficulties with autocorrelations among individual energy estimates.
All calculations were performed on single workstations, though
extensions to large systems would require a parallel implementation.

The various energy
estimators are plotted as a function of the size of the time step
$\Delta\tau$ in Fig.~\ref{figj}. The exact ground-state energy is 
shown as a
circle at the extrapolated $\Delta\tau = 0$ limit.  The two dashed
curves illustrate the growth and mixed estimates $E_g$ and $E_m$.
$E_g$ is simply obtained from the change in overlap with
iteration
\begin{equation} 
e^{- \Delta\tau E_g } =
\frac{\langle \Psi_T | \Psi_c (\tau+\Delta\tau) \rangle}
{\langle \Psi_T | \Psi_c (\tau) \rangle} \ ,
\end{equation} 
while $E_m$ is obtained by direct
evaluation of Eq. \ref{eq:eestimates}.  Since the propagator is
approximate, these two estimates coincide only in the limit of small
$\Delta\tau$.

As apparent from the figure, these two estimates lie slightly below
the exact energy.  The value of $E_g$, extrapolated to $\Delta\tau=0$,
is $-17.517(2)$. During the course of this calculation, we also
evaluated $E_d$.  In this case, $E_d$ is small
and positive, and adding this difference to $E_g$ should produce a
variational upper bound to the ground-state energy. Extrapolating to
$\Delta\tau = 0$, we find $E_d = 0.010(1)$, and hence $E_v =
-17.506(2)$. The accuracy of the variational bound is quite
surprising: the exact energy is recovered with an accuracy of two
parts in $10^4$, or better than 99.9$\%$ of the difference between the
exact and trial state energies.

We also plotted in Fig.~\ref{figj} the result of a direct calculation 
of $E_v$. Since we have independent calculations of the left and 
right-hand
states, it is possible to combine these into a direct calculation of
the variational energy $E_v$.  In contrast to the other estimators,
this calculation should yield a variational upper bound independent of
the time step $\Delta\tau$.  We find this to be true, but with a
somewhat larger statistical error than the other estimators.  There
also appears to be some residual statistical bias resulting from the finite
population size. This estimator may be more difficult to compute
reliably for larger system sizes.

The statistical errors for $E_d$ also increase rapidly with system 
size.  Primarily this is a result of a large statistical error in the 
denominator (Eq. \ref{eq:evdiff}). Particularly for our simple choices 
of trial states, the overlaps of the
configurations representing the left- and right-hand population can
vary dramatically.  The alternative method of altering the constraint
slightly (Eq. \ref{eq:alterconstraint}) should produce a more
favorable scaling with system size, though it remains to be
demonstrated that 
this is practical for very large simulations.

For the 4$\times$4 case, we explicitly changed the constraint by 
introducing a finite value of $\alpha$. We could achieve a variational 
upper bound by setting $\alpha = 0.0005$; for $\alpha = 0.0003$ we 
could explicitly see that the sign of the overlap $\langle \Psi_c | 
\Psi_d \rangle$ could lead to a violation of the upper bound. For
$\alpha = 0.0005$  
and a time step of 0.005,
we obtain a mixed estimate $E_m  = -17.518(3)$ and a growth
estimate $E_g = -17.505(3)$.  Recall that only $E_g$ provides an
upper bound in the limit of zero time step. However, the small
difference $E_m - E_g$ indicates the accuracy of the solution.
Extrapolating to zero time step yields $E_g = - 17.510(10)$.

We also considered a 6$\times$6 lattice with 13 spin up and 13 spin
down electrons, again for $U$=8. We are unaware of any exact or QMC
calculations for this system size at this filling.  These larger
system size results are meant to serve as guides for future use rather
than exhaustive calculations. They were obtained on
single-cpu workstation over the course of a few days.

For the $6\times 6$ system the constant $\alpha$ must be decreased
significantly.  This is rather natural as one would expect it to
scale roughly with a small power of the number of single-particle
orbitals.  Again we use approximately 1000 configurations
averaged over 30-100 individual blocks with a total propagation time 
of 2-4 per block. 
Here it is not clear if the original choice of
$\alpha = 0$ provides a variational upper bound, estimates of
$E_d$ bracket zero within the statistical errors of the calculation.
For $\alpha = 0$ we obtain $E_m = E_g = -36.05 (05)$.

Increasing the constant $\alpha$ to $10^{-6}$ provides a variational
upper bound.  In this case we obtained $E_m = -35.75(05)$ and
$E_g = -34.55(10)$ for a time step of $\Delta \tau = 0.005$.
Extrapolating to $\Delta \tau = 0$ yields $E_m = -35.80(05)$ and
$E_g = -35.25(20)$.  It is possible that this bound could be
further improved by using a somewhat smaller value of $\alpha$.
Again, the few per cent difference between $E_g$ and $E_m$
indicates the accuracy of the calculation.

\section{Concluding Remarks}
\label{sec5}

We presented several differences between the constrained-path and
fixed-node Monte Carlo methods, some major and some 
minor. The most significant consequence of these differences is the
mixed estimator in the CPMC method not being an upper bound to the
exact energy as it is in the fixed-node method.  Alternate ways of
producing an upper bound have been introduced. 

While not an upper bound, the mixed estimator in the CPMC method was
argued to be very near the exact answer. 
Experience shows it is almost always above
the exact answer, and in cases where the CPMC results have been
compared to fixed-node results, the CPMC mixed estimates of the energy
always lie closer to the exact answer than the upper bound
produced by the fixed-node method.\cite{FNexamples} Presumably this
accuracy is a consequence of the quality of the estimates of the wave
function. As a rule of thumb, we find that the fewer nodal crossings
the more accurate is the prediction of the energy. This observation is
supported by the method discussed in Section~IV to correct the mixed
estimate so it produces an upper bound. This method depends on 
computing a contribution from those walkers thrown away; if none are 
thrown away (after the sampling is from the limiting distribution), 
then the CPMC method in fact becomes exact.

There are several other differences between the methods worth
mentioning. In one continuous spatial dimension, coincident 
planes ($x_i = x_j$ for all $i$ and $j$ corresponding to the same spin 
species) exhaust the nodal surface set for local potentials $V$; 
therefore, in the fixed-node method one can get the exact ground-state 
energy by using any nodal surface with this property. The free-fermion 
wave function suffices. For general lattice fermion problems, even in 
one spatial dimension, the situation is more complicated: there are 
extra nodal surfaces which are not coincident planes. For certain 
classes of Hubbard-like models the latter exhaust the whole nodal set; 
therefore it is possible to avoid the sign problem and get the exact 
solution.\cite{lee} On the other hand, the lattice version of the 
fixed-node method \cite{FNlatt} always provides a variational upper 
bound to the exact ground-state energy regardless of the 
dimensionality, unless the constraining state is the exact 
ground-state wave function. We have been
unable to develop a similar understanding for the constrained-path
method where in one dimension we observe an absence of nodal crossings
and the mixed estimate of the energy agreeing with exact results to
statistical accuracy.\cite{torsten} We comment that care must be taken 
in using nodal crossing rule of thumb. The accuracy of the Trotter
approximation is controlled by the size of $\Delta\tau$. If
$\Delta\tau$ is large, the approximation is poor, and nodal crossings
can be induced into a problem for which there is no sign problem. If
$\Delta\tau$ is too small, the propagation through phase space is too
slow. For a poor choice of importance sampling population control can 
sweep away walkers that should cross the surface before they actually 
do.

We also remark that in the CPMC method there is much less
need to perform a variational optimization of $|\Psi_T\rangle$
through Jastrow or Gutzwiller factors as seems to be necessary in the
fixed-node method.\cite{mariana2} This optimization process does not
affect the nodal surface but does reduce the energy of the starting
configuration. While there is always some advantage in doing this, the
results of the CPMC method display considerable robustness to the
choice of the constraining wave function which is also typically used
as the starting configuration. Simple choices, like free-fermion wave
functions, seem to work well. Quite different choices of 
$|\Psi_T\rangle$ usually give satisfyingly similar 
results.\cite{mariana2}

In closing, we remark that some of our observations about the mixed
estimator for the energy might be useful in constructing and assessing
estimation procedures used in the standard auxiliary-field projector
quantum Monte Carlo (AFQMC) method.\cite{koonin,sorella,white} In that
method, the energy or more generally some observable ${\cal O}$ is
typically estimated\cite{loh,fahy,sorella} from
\begin{equation}
 {\cal O}(\tau,\tau')
   =\frac{\langle\Psi_T|e^{-(\tau-\tau')H}{\cal O}e^{-\tau'H}|\Psi_T
\rangle}{\langle\Psi_T|e^{-(\tau-\tau')H}e^{-\tau'H}|\Psi_T\rangle} \ ,
\end{equation}
and then for large $\tau$ this expression is either averaged over
several values of $\tau'$ or evaluated at just
$\tau'=\tau/2$. Clearly, the latter procedure may be preferable, even
though the former may have lower variance, as this estimator can be
rewritten as
\begin{equation}
 {\cal O}(\tau,\tau')=\frac{\langle\Psi_L|{\cal O}|\Psi_R\rangle}
                    {\langle\Psi_L|\Psi_R\rangle} 
\end{equation}
revealing that for $\tau'\not=\tau$ it is basically just a mixed
estimator. For estimating the energy or an observable that commutes
with $H$, the utility of this estimator depends on how close either
$\langle\Psi_L|$ or $|\Psi_R\rangle$ approaches the ground state wave
function. For estimation of observables that do not commute with the
Hamiltonian $H$, its utility depends on how close both
$\langle\Psi_L|$ and $|\Psi_R\rangle$ approach the ground state wave
function.

\section{Acknowledgments}

We gratefully acknowledge illuminating discussions with Erik Koch 
on the variational aspects of the constrained-path method. 
Most of this work was supported by the Department of Energy. S.Z. is
supported by the National Science Foundation under grant DMR-9734041
and the Research Corporation.

\appendix\section*{Illustrative Example}

We now introduce an exactly solvable fermion model to illustrate and
visualize several features of the constrained-path method. This model
has the Hamiltonian
\begin{equation}
H = -t \left( c_1^{\dagger} c_2^{\;} + c_2^{\dagger} c_3^{\;} + 
c_2^{\dagger} c_1^{\;} + c_3^{\dagger} c_2^{\;} \right) + 
U \left( n_1 n_2 + n_2 n_3 \right) = T + V \ ,
\end{equation}
and corresponds to spinless fermions coupled through a nearest 
neighbor repulsive interaction $U$ on a three site lattice with open boundary 
conditions. In the following we will concentrate on the two-particle 
solutions for which the ground state is
\begin{equation}
| \Psi_0 \rangle = \frac{1}{\sqrt{2 t^2 + (U - E_0)^2}}  \left[ 
t\ ( c_1^{\dagger} c_2^{\dagger} +  c_2^{\dagger} c_3^{\dagger} )
+ (U - E_0) \ c_1^{\dagger} c_3^{\dagger} \right] | 0 \rangle \ ,
\end{equation}
and $E_0$, the ground-state energy, equals $U/2 - \sqrt{(U/2)^2 + 2
t^2}$.

Since we want to study explicitly the (imaginary-time) evolution of
the distribution of Slater determinants which arise as a consequence
of the constrained-path approach, we need to have some way of
parameterizing the differentiable manifold of Slater determinants of
two particles that has dimension $N(M-N)= 2(3-2) = 2$. We can
parameterize a state in the two-particle Hilbert space ${\cal H}_2$
that belongs to the set of {\it normalized} Slater determinants by the
$3 \times 2$ matrix
\begin{equation}
\Phi = \begin{pmatrix}
       \cos (\theta_1 -\theta_2) & 0                 \\
       \cos \theta_1             & \cos \theta_2     \\
       \sin \theta_1             & \sin \theta_2  
       \end{pmatrix}
\ . 
\end{equation}
Therefore, the two angles $(\theta_1, \theta_2)$ specify a point in 
the manifold, and any state $| \Psi \rangle$ belonging to the Hilbert 
space ${\cal H}_2$ having support in that manifold can be represented 
by $\Psi[\theta_1, \theta_2] \equiv \langle \phi | \Psi \rangle$.  
In general, the usual property of a fermion wave function to be 
totally antisymmetric in spin-coordinate space is lost in this 
new representation. For instance, the ground state of our model 
fermion system is given by
\begin{equation}
\Psi_0[\theta_1, \theta_2] = \frac{1}{\sqrt{2 t^2 + (U - E_0)^2}} 
\left[ \cos(\theta_1 - \theta_2) \ \left(t \cos \theta_2 + (U-E_0) 
\sin \theta_2 \right) + t \sin(\theta_2 - \theta_1)\right] \ ,
\end{equation}
which is neither symmetric nor anti-symmetric under permutations of
$\theta_1$ and  
$\theta_2$. Contour plots of it can be found in Fig.~\ref{fig2} for 
different values of the interaction strength $U$. In the 
non-interacting case, $U = 0$, the ground state is a single Slater 
determinant represented by the matrix
\begin{equation}
\Phi_T = \begin{pmatrix}
\sqrt{\frac{3}{4}}                            & 0                   \\
\frac{1}{2} \left( 1+ \sqrt{\frac{2}{3}}\right) & \sqrt{\frac{1}{3}}\\
\frac{1}{2} \left( \sqrt{2}-\sqrt{\frac{1}{3}} \right) & 
\sqrt{\frac{2}{3}}
       \end{pmatrix}
\ , 
\end{equation}
corresponding to the point $\theta_{1,T}=\theta_{2,T} - \pi/6$, 
$\theta_{2,T}= \arccos 1/\sqrt{3}$ in the manifold.

For a given value of $U$ we would like to project out the ground state 
from this non-interacting state. For 
simplicity, we will leave out importance sampling and use a first
order Trotter decomposition for the  short time propagator
\begin{equation}
e^{-\Delta\tau H} | \phi \rangle = e^{-\Delta\tau V} e^{-\Delta\tau T} 
| \phi \rangle  +  {\cal O}(\Delta\tau^2) \ ,
\label{prop}
\end{equation}
and stochastically iterate this expression for each walker. In a 
matrix representation this iteration is equivalent to a product of 
non-commuting random matrices, i.e. $\Phi(\tau) = {\cal U}(\tau) 
\Phi_T = \prod_{i=1}^{n} \left( \exp[-\Delta\tau V] \exp[-\Delta\tau 
T] \right) \Phi_T$, where the symmetric matrix $\exp[-\Delta\tau T]$ 
is given by
\begin{equation}
e^{-\Delta\tau T} = \begin{pmatrix}
\cfrac{u+1}{2} & \cfrac{v}{\sqrt{2}} & \cfrac{u-1}{2} \\
\cfrac{v}{\sqrt{2}} & u             & \cfrac{v}{\sqrt{2}} \\
\cfrac{u-1}{2} & \cfrac{v}{\sqrt{2}} & \cfrac{u+1}{2} 
                    \end{pmatrix} 
\end{equation}
with $u=\cosh(\sqrt{2} \ t \Delta\tau)$, $v^2 = u^2 -1$. After the use 
of the discrete ($U > 0$) Hubbard-Stratonovich transformation  
$\exp[-\Delta\tau U n_j n_{j+1}] = \frac{1}{2} \exp[-\Delta\tau U 
(n_j+n_{j+1})/2] \sum_{x=\pm 1} \exp[\mu x (n_j - n_{j+1})]$, the
interaction part of the propagator at any imaginary-time slice $i$ 
is represented by one of the 4 diagonal random matrices (each chosen 
with probability 1/4, i.e. $P({\bf x})=1/4$)
\begin{equation}
e^{-\Delta\tau V(i)} \equiv \begin{pmatrix}
                            \alpha_i& 0  & 0 \\
                             0 & \beta_i & 0 \\
                             0 & 0 & \gamma_i
                            \end{pmatrix}
= e^{-\Delta\tau U/2} \left\{ \begin{pmatrix}
e^{\pm \mu}     & 0                    & 0 \\
0               & e^{-\Delta\tau U/2}  & 0 \\
0               & 0                    & e^{\mp \mu}
                                  \end{pmatrix} \ , \
                                  \begin{pmatrix}
e^{\pm \mu}     & 0                    & 0 \\
0               & e^{-\Delta\tau U/2 \mp 2 \mu}  & 0 \\
0               & 0                    & e^{\pm \mu}
                                  \end{pmatrix}  \right\} \ ,
\end{equation}
where $\cosh \mu = \exp[\Delta\tau U/2]$. Although 
${\cal U}(\tau)$ is a random matrix, its determinant $\det {\cal 
U}(\tau) = \exp[-2 \tau U]$ is not a random number.

After a short propagation (Eq.~\ref{prop}), each point of the Slater
determinant manifold, representing the state of the system, performs a
Brownian walk in $\theta-$space. Therefore, one can consider
$\theta_1(\tau)$ and $\theta_2(\tau)$ (such that
$\theta_1(0)=\theta_{1,T}$, and $\theta_2(0)=\theta_{2,T}$) as a set
of random variables defining a random walker in imaginary
time. However, in the CPMC method, to avoid the fermion sign
problem, we constrain the walker with a constraining state which here
we choose it to be $|\Psi_0(U=0)\rangle$. In other words, each time
Eq.~\ref{prop} is iterated, we ask, Has the sign of $\det \left[
\Phi^{\rm T}_T\Phi(\tau) \right]$ changed ?  If it has, then we kill the 
walker.

For the present example, the determinant can be computed analytically:
At any time step $n \ge 1$
\begin{equation}
\det \left[ \Phi^{\rm T}_T \Phi(\tau) \right] = \frac{u+v}{2^{n+1}} 
\left[ \alpha_1 \beta_1 \ f_2 + \beta_1 \gamma_1 \ g_2 + 2 \alpha_1 
\gamma_1 \ h_2 \right] \ , 
\label{deter}
\end{equation}
where the functions $f_i$, $g_i$ and $h_i$ are most simply obtained
from the backwards recursion relations
\begin{eqnarray}
f_i &=& \alpha_i \beta_i \ (u+1) \ f_{i+1}  + \beta_i \gamma_i \ (u-1) 
\ g_{i+1} + 2 \alpha_i \gamma_i \ v \ h_{i+1} \nonumber \\
g_i &=& \alpha_i \beta_i \ (u-1) \ f_{i+1}  + \beta_i \gamma_i \ (u+1) 
\ g_{i+1} + 2 \alpha_i \gamma_i \ v \ h_{i+1} \\
h_i &=& \alpha_i \beta_i \ v \ f_{i+1}  + \beta_i \gamma_i \ v \ 
g_{i+1} + 2 \alpha_i \gamma_i \ u \ h_{i+1} \nonumber
\end{eqnarray}
with $f_{n+1} = g_{n+1} = h_{n+1} = 1$ being the initial conditions.

Clearly the determinant Eq.~\ref{deter} is always positive, 
independent of the values of $n$, $\Delta \tau$ and $U$, which 
means that there is no sign problem. Nevertheless, this example helps 
to illustrate several important issues. The first is that the 
exact ground-state energy can be stochastically obtained (as we will 
show below) even if the nodal surface of $\langle \phi | 
\Psi_T \rangle$ is approximate. The second issue deals with the 
practical numerical implementation of the method. A ``false'' sign 
problem (i.e. $\det \left[ \Phi^{\rm T}_T \Phi(\tau) \right] < 0$) can 
occur as a consequence of numerical round-off errors. In fact, the use 
of matrix stabilization techniques is crucial to avoid such 
phenomenon.\cite{loh}

Figure~\ref{fig3} shows the energy mixed estimator $E_m$ as a function 
of imaginary time. All walkers start ($\tau=0$) at the point 
$(\theta_{1,T},\theta_{2,T})$ in $\theta$-space. That means that 
$E_m(\tau = 0)= U/2 -\sqrt{2} t$, and as time evolves, $E_m 
\rightarrow E_0$ if the constraint is not evoked. As mentioned above 
this is the case if matrix stabilization techniques are properly used, 
otherwise we observe ``false'' nodal crossings. A typical 
random walker $\theta(\tau)$ is displayed in Fig.~\ref{fig4}. This 
figure clearly denotes that most of the time the walker prefers to 
stay near the upper right corner of the space. This behavior is 
evidenced in the lower panel of the same figure, where it is shown 
that when the walker is at the upper right corner in $\theta$-space 
the overlap with the exact ground state is maximum (in fact, it is 
almost one).

\newpage
% figures follow here
%
% Here is an example of the general form of a figure:
% Fill in the caption in the braces of the \caption{} command. Put the label
% that you will use with \ref{} command in the braces of the \label{} command.
%
 \begin{figure}
 \caption{Contour plots of the two-fermion ground-state wave function 
$\langle \phi | \Psi_0 \rangle$ for various values of the interaction 
strength $U/t$. Notice the differences among the nodal structures.}
 \label{fig1}
 \end{figure}

 \begin{figure}
 \caption{Energy estimators as a function of time step for the  
4$\times$4 Hubbard model described in the text.  Solid symbols indicate
estimators which are variational upper bounds to the exact energy.}
 \label{figj}
 \end{figure}

 \begin{figure}
 \caption{Contour plots of the two-fermion ground-state wave function 
$\langle \phi | \Psi_0 \rangle$ for various values of the interaction 
strength $U/t$. }
 \label{fig2}
 \end{figure}

 \begin{figure}
 \caption{Energy mixed estimator as a function of imaginary time 
averaged over $N_w$ walkers. The horizontal dashed line indicates 
the value of the exact ground-state energy $E_0=-0.19615 t$.}
 \label{fig3}
 \end{figure}

 \begin{figure}
 \caption{The upper panel shows a typical random walk in 
$\theta$-space. At $\tau$=0 the walker starts at 
$(\theta_{1,T},\theta_{2,T})$. The random walk never crosses the
nodal surface of $\langle \phi | \Psi_T \rangle$. The lower panel 
displays the overlap of the walker with the exact ground state, and 
the distance of the walker from the origin. Notice the clear 
correlation between these two quantities.}
 \label{fig4}
 \end{figure}
% tables follow here
%
% Here is an example of the general form of a table:
% Fill in the caption in the braces of the \caption{} command. Put the label
% that you will use with \ref{} command in the braces of the \label{} command.
% Insert the column specifiers (l, r, c, d, etc.) in the empty braces of the
% \begin{tabular}{} command.
%
% \begin{table}
% \caption{}
% \label{}
% \begin{tabular}{}
% \end{tabular}
% \end{table}


\begin{references}
\bibitem{example} Other examples were of course computed. For example,
for the same system but with $U=20t$, the exact answer is -15.452
while the CPMC result is -15.460(3). This suggests that the violation
of the upper bound does not obviously get worse as $U$ increases. On
the other hand, we know the violation cannot get worse as $U$
decreases.
\end{references}
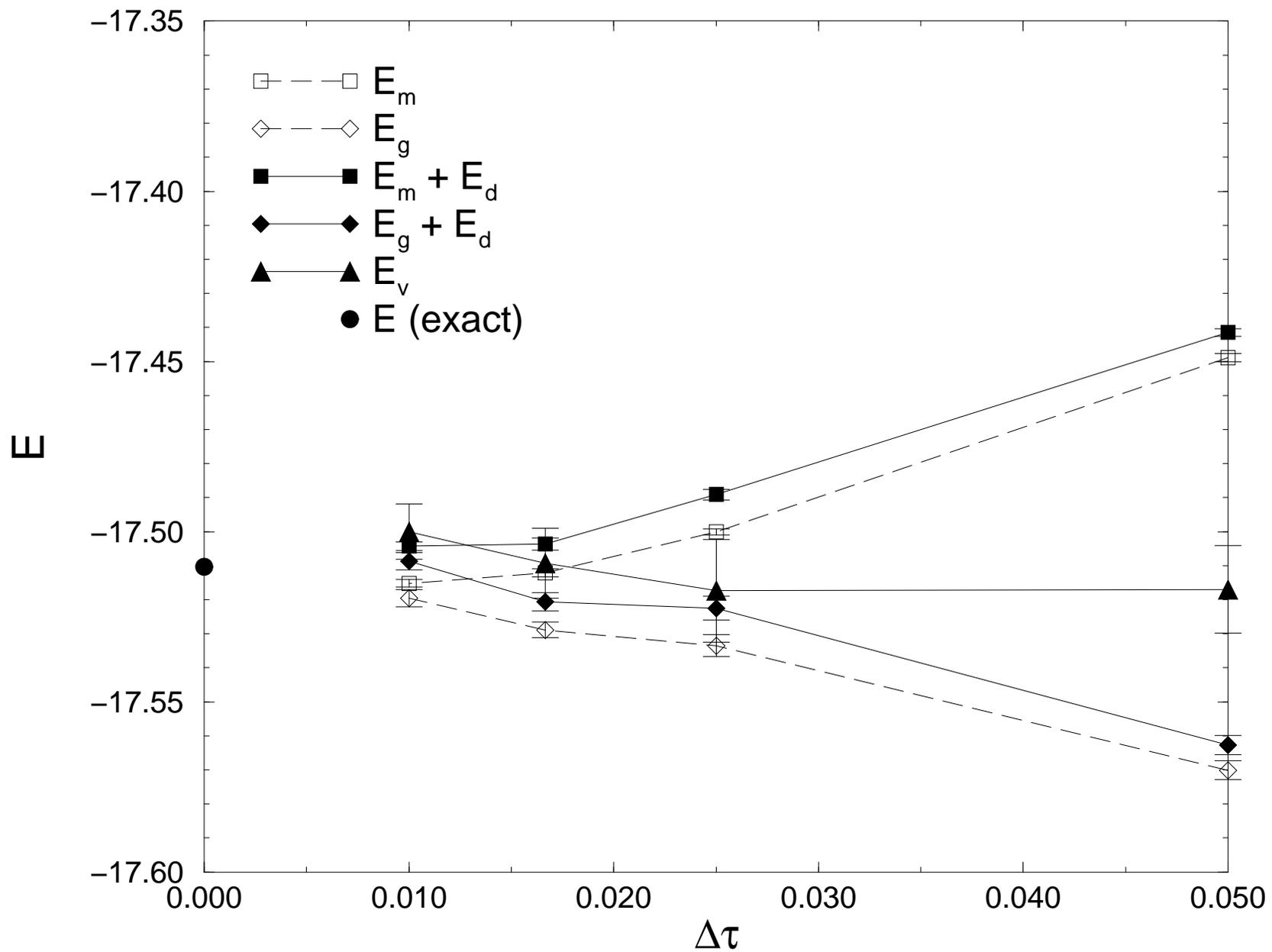
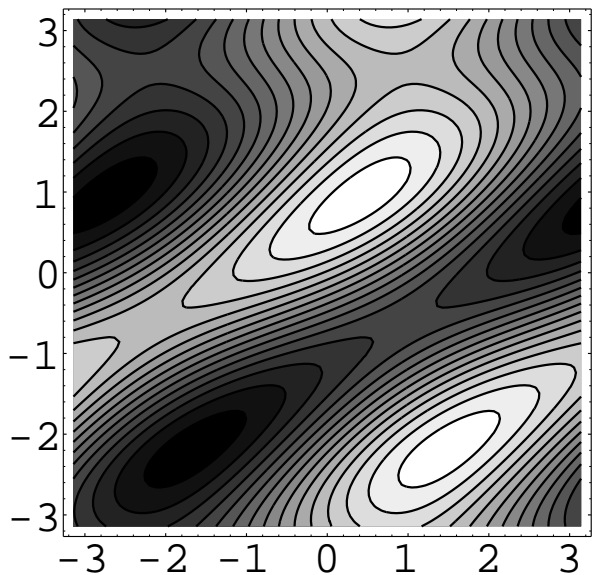
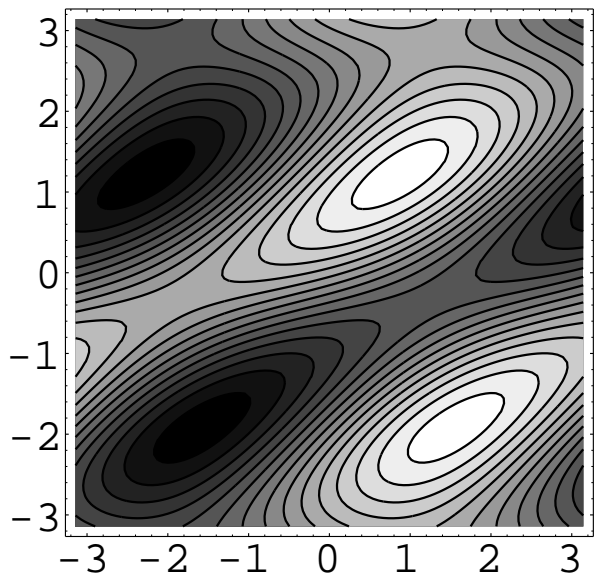
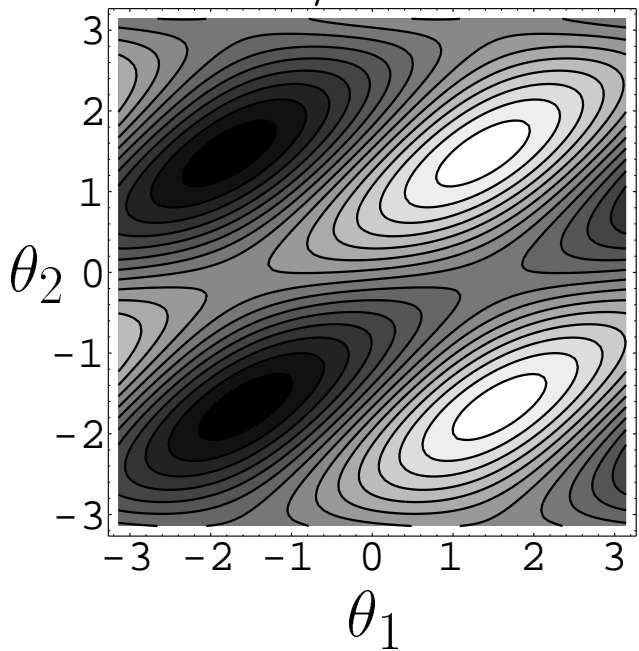
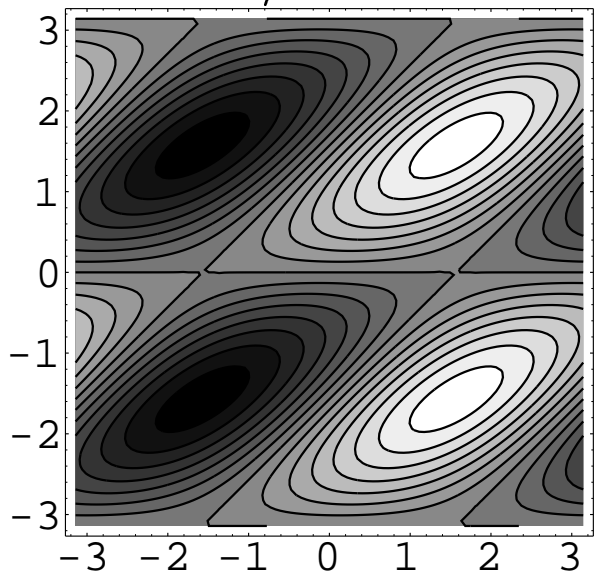
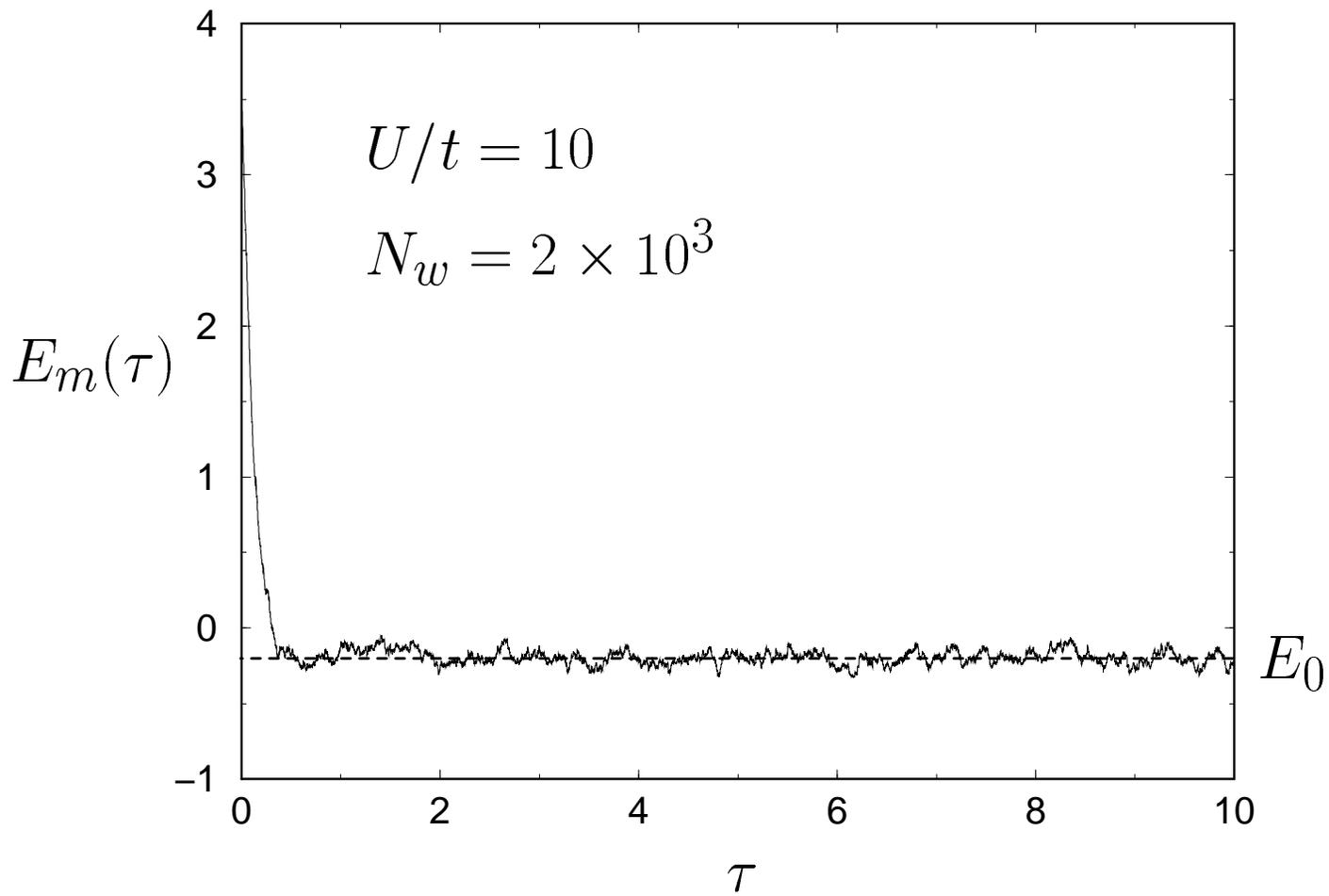
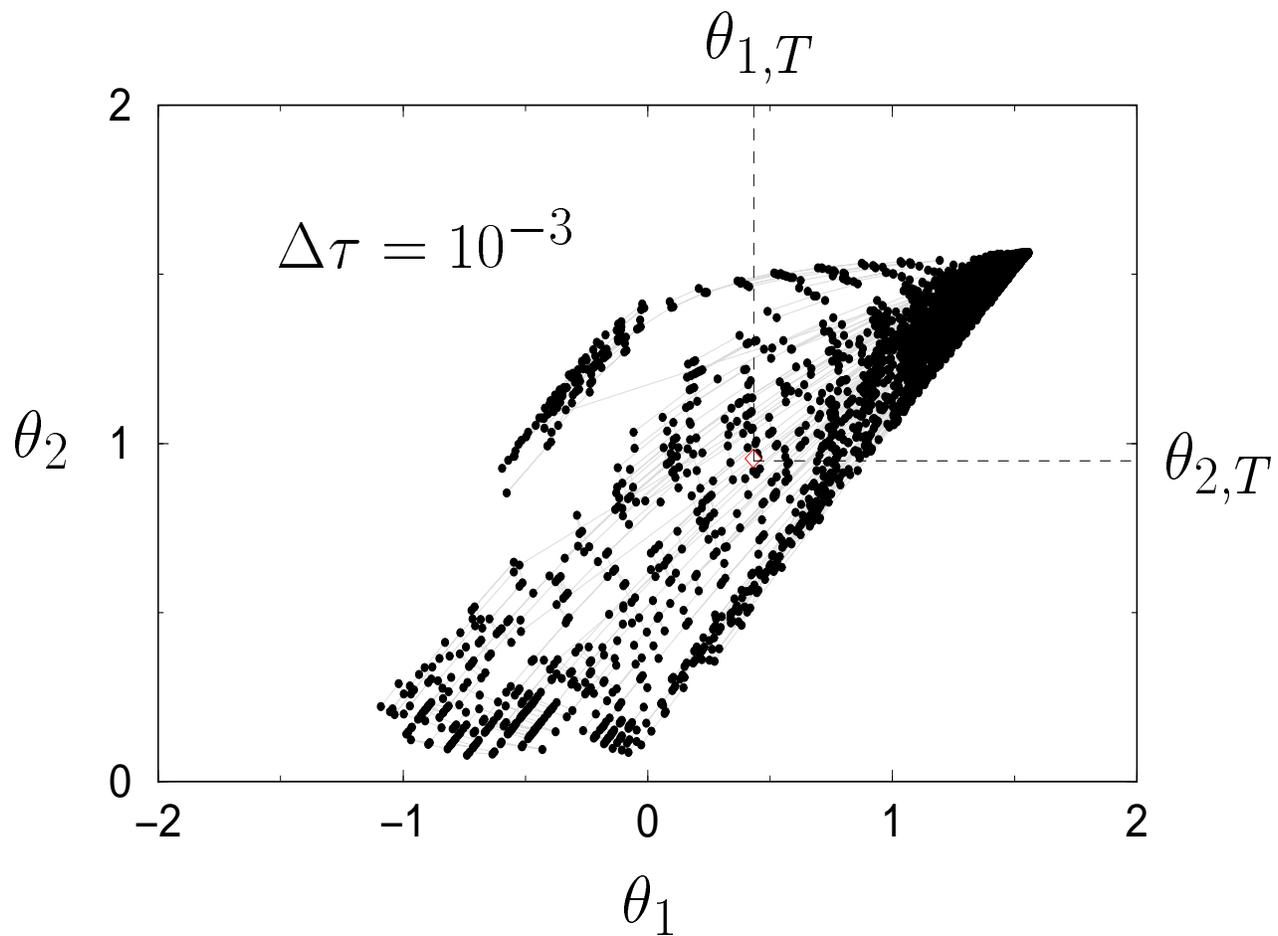
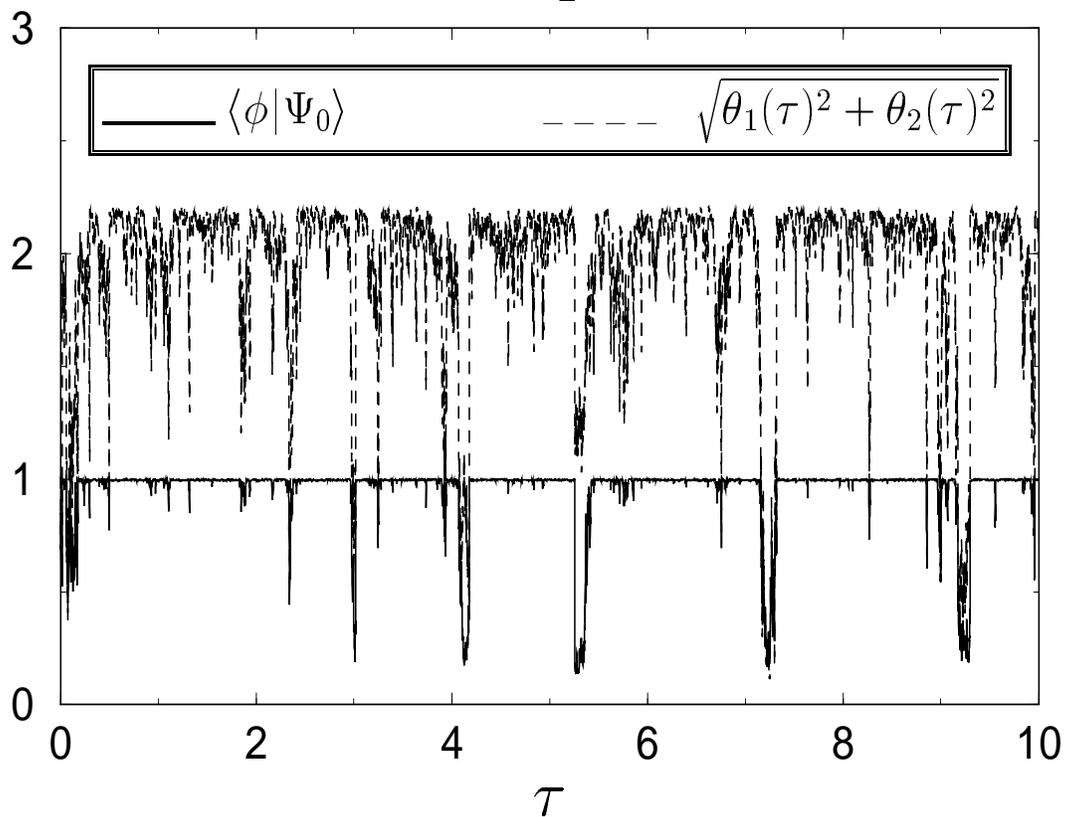
\end{document}